# A Relational Approach to Functional Decomposition of Logic Circuits

## Tony, T. Lee [a],[*], Tong Ye [a]

[a]*Department of Information Engineering, The Chinese University of Hong Kong, Shatin, N.T., Hong Kong, China*

**Abstract**

Functional decomposition of logic circuits has profound influence on all quality aspects of the cost-effective implementation of modern digital systems. In this paper, a relational approach to the decomposition of logic circuits is proposed. This approach is parallel to the normalization of relational databases, they are governed by the same concepts of functional dependency (FD) and multi-valued dependency (MVD). It is manifest that the functional decomposition of switching function actually exploits the same idea and serves a similar purpose as database normalization. Partitions play an important role in the decomposition. The interdependency of two partitions can be represented by a bipartite graph. We demonstrate that both FD and MVD can be represented by bipartite graphs with specific topological properties, which are delineated by partitions of minterms. It follows that our algorithms are procedures of constructing those specific bipartite graphs of interest to meet the information-lossless criteria of functional decomposition.

## 1  Introduction

Functional decomposition of logic circuits has profound influence on all quality aspects of the cost-effective implementation of modern digital systems [1,2]. It is clear that the single table specification of a large size switching function may contain undesirable redundancies, which can only be eliminated by factorizing the function into several small and independent functions. It is evident that the decomposition of a single table into smaller tables not only can reduce the complexity, but it can also increase the scalability, and improve the reliability of the logic circuits. A general information lossless decomposition theory of relational information systems has been proposed in [3]. This paper addresses the specific application of this theory to the decomposition of logic circuits.

---

* Corresponding author.
  *Email address:* `ttlee@ie.cuhk.edu.hk` (Tony, T. Lee).

Various approaches [4] have been proposed to explore the functional decomposition of switching function. Earlier works in this area were initiated by R. L. Ashenhurst [5] and H. A. Curtis [1]. Their approach later has been widely adopted by many other researchers [6–10] and extensively applied to Field Programmable Gate Arrays (FPGAs) [11, 12], multiple-valued logics [13–16], machine learning [17, 18], and Very Large Scale Integration (VLSI) [19].

In this paper, we propose a relational approach deduced from the decomposition theory developed in relational databases [20, 21]. The relational data model is the simplest but most versatile way to manage information. It is focused on the organization of data into collections of tables, called *relations*. In relational databases, normalization is the process of removing redundant data from relational tables by decomposing (splitting) a relational table into smaller tables by projection, which in turn saves space and reduces manipulation anomalies. In order to be correct, decomposition must be information-lossless. That is, the new tables can be recombined to recreate the original table without creating any spurious data. Considering the truth table of a switching function as a relation, it naturally leads us to construct functional decomposition algorithms based on the intrinsic structure properties of tables unfolded in the course of developing the normalization theory of relational databases.

There are two interwind themes in this paper. First, we introduce the notions of relational data model that provide us with the table structures that we need to study the decomposition of switching function. Our investigation reveals that the theory of functional decomposition of logic circuits is parallel to the normalization theory of relational databases, they are governed by the same concepts of functional dependency (FD) and multi-valued dependency (MVD) [21]. It is manifest that the decomposition of switching function exploits the same idea and serves a similar purpose as database normalization.

Our second theme is to utilize concepts of FD and MVD in the construction of algorithms for decomposition of switching functions. Partitions play an important role in the information lossless decomposition of relational systems. Hartmanis and Stearns [22] investigated the serial and parallel decompositions of finite state machines by using the partitions with substitution property introduced by Birkhoff [23]. In logic circuit decomposition, the set of minterms of a truth table is partitioned into blocks according to some equivalence relations, and each block, or equivalence class, of the partition represents a set of minterms of a component circuit. Following the theory developed by Birkhoff [23], partitions corresponding to the component subsystems must possess the commutative property. Rota gives an interesting information-theoretic interpretation of commutative partitions in [24]. The commutative property is essentially a general algebraic formulation of independency of two partitions. We show that partitions of minterms associated with both FD and MVD can be represented by bipartite graphs with specific topological properties. It follows that our algorithms are procedures of synthesizing those specific bipartite graphs of interest to meet the lossless criteria of functional decomposition.

The principle of our basic disjoint decomposition algorithm lies on the merge of equivalent columns in the decomposition chart to remove redundancies in the underlying truth table. The bipartite graphic representations of FDs and MVDs are facilitated to prove that the information contained in the truth table is preserved by the decomposition algorithm. The decomposition chart of non-disjoint decomposition can be organized in diagonal form, such that each sub-chart in the diagonal can be treated as a disjoint decomposition chart. Thus, it is straightforward to generalize the basic algorithm to the case of non-disjoint decomposition. The decompositions of incompletely specified functions are governed by the same set of rules, except that there are many alternative combinations to merge the columns of the decomposition chart. Due to unspecified values in the truth table, the final solution may not be unique, and it is determined by cliques of compatible columns. Our relational approach is developed in the context of partitions associated with logic variables in the truth table. Thus, the proposed decomposition algorithms can be naturally extended to multi-valued logic circuits without any amendments.

The sequel of this paper is organized as follows. Section 2 provides the motivation of relational approach to the decomposition of switching function. Section 3 introduces notions of relational data model to facilitate necessary background information. In Section 4, we construct the algorithm for simple disjoint decomposition. In Section 5, the algorithm is generalized to multiple decomposition. Section 6 considers the non-disjoint decomposition. In Section 7, we discuss the incompletely specified function. Section 8 is devoted to the decomposition of multi-valued logic circuits. Finally, the conclusion and future research direction are summarized in Section 9.

## 2    Ashenhurst-Curtis Decomposition

To motivate our relational approach of functional decomposition, we will closely examine an example in this section to demonstrate that Ashenhurst-Curtis decomposition of switching function is essentially a sequence of relational operations performed on the truth table of the underlying logic circuit.

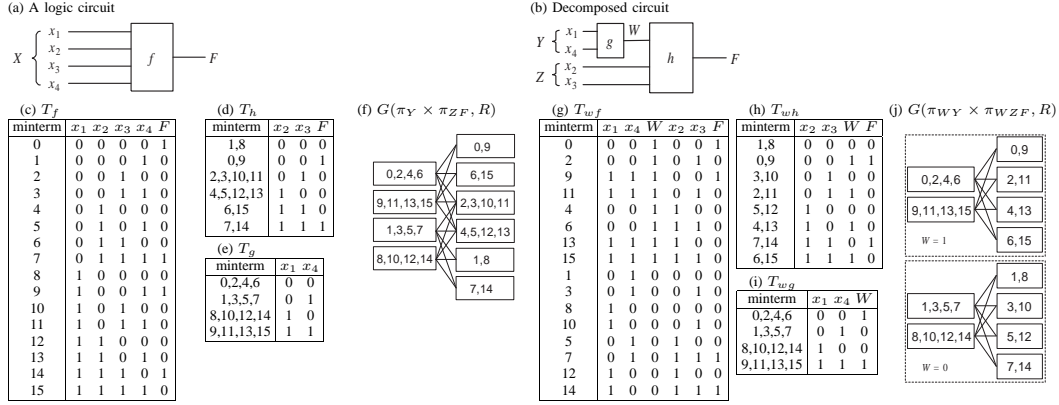

Fig. 1. An example of logic circuit decomposition.

**Example 1** *As shown in Fig.1(a) and (b), the switching function $f$ on the set of variables $X$, $F = f(X)$, is decomposed into two functions $W = g(Y)$ and $F = h(W,Z)$, where $Y$ and $Z$, called* bound set *and* free set *in [1], respectively, are subsets of $X$ such that $Y \cup Z = X$.*

*The truth table $T_f$ of the switching function $f(X)$ on the set of input variables $X = \{x_1, x_2, x_3, x_4\}$ is shown in Fig.1(c), in which the bridge variable $W$ in the decomposed function shown in Fig.1(b) does not exist and its values are unknown. The Ashenhurst-Curtis decomposition can be considered as a procedure to determine the values of this unknown bridge variable $W$ for given bound set $Y = \{x_1, x_4\}$ and free set $Z = \{x_2, x_3\}$.*

*The logic circuit shown in the schematic diagram Fig.1(b) will break down into two components if the link representing the bridge variable $W$ is removed. The truth tables $T_h$ and $T_g$ indicating the states of those two components are given in Fig.1(d) and (e), which are projections of the truth table $T_f$ on the domains $x_2 x_3 F$ and $x_1 x_4$, respectively. The minterms in the same row of $T_h$ or $T_g$ possess the same value in their corresponding domains.*

*Inspecting the three truth tables $T_f$, $T_h$ and $T_g$, it is easy to realize that each minterm in $T_f$ is the intersection of a set of minterms corresponding to a row in $T_h$ and a set of minterms in $T_g$. This relationship is expressed explicitly by the bipartite graph depicted in Fig.1(f), in which each node represents a set of minterms of $T_h$ and $T_g$, and each edge indicates a minterm of $T_f$.*

*The key to decompose the logic circuit lies on the determination of the values of the bridge variable $W$ for each minterm. In this example, assigning the value of $W$, determined by the Ashenhurst-Curtis decomposition procedure, to each minterm in the truth table $T_f$, we obtain the the truth table $T_{wf}$ of the decomposed circuit $f$. It follows that the truth tables of component circuits $h$ and $g$, given in Fig.1(h) and (i), are projections of $T_{wf}$ on $x_2 x_3 WF$ and $x_1 x_4 W$, respectively.*

In Section 3, we will show that the bipartite graph in Fig.1(j) actually represents a multi-valued dependency (MVD) that ensures the lossless decomposition of the truth table $T_{wf}$. Comparing the two bipartite graphs given in Fig.1(f) and (j), we observe that the nodes of the bipartite graph in Fig.1(j) are subsets of those nodes in the bipartite graph displayed in Fig.1(f), and this "node splitting" is completely determined by the values of $W$. That is, all the minterms in each connected component of the bipartite graph in Fig.1(j) have the same $W$ value.

Owing to the one-to-one correspondence, illustrated in the proceeding example, between the values of bridge variable $W$ and the connected components of the bipartite graph, the proposed decomposition algorithm will synthesize the bipartite graph of the decomposed circuit based on conditions of FDs and MVDs first, and then values of the bridge variable $W$ can be determined immediately.

In the next section, we provide the basic notions on relational data model and properties of FDs and MVDs to facilitate information that are necessary for synthesizing the disjoint decomposition algorithm detailed in section 4. Subsequently, we show that this procedure can be further generalized to multiple decomposition, non-disjoint decomposition, incomplete specified function and multi-valued logic circuit in a straightforward manner.

## 3 Relational Model of Logic Circuits

Originally, the relational model was developed for databases [20], the organization of data into collection of tables. Consider the truth table of logic circuit as a relation and take account of structure issues arising in relational databases, it is natural to explore functional decomposition of logic circuits on the basis of functional dependency (FD) and multi-valued dependency (MVD) [21], which are vital concepts to study the information-lossless decomposition of relational tables.

### 3.1 Relational model

In relational model, a relation is a two-dimensional table, each row in the table is called a *tuple*, and the columns of the table are given names, called *attributes*. As an example, the truth table shown in Fig.1(c) is a relation $R[\Omega]$ with the set of attributes $\Omega = \{x_1, x_2, x_3, x_4, F\}$, and each minterm is a tuple of $R[\Omega]$. Throughout the discussions in the sequel, we will use the term tuple to indicate minterm, and only show the indices of those tuples in all figures.

Let $X$, $Y$, and $Z$ be the sets of attributes, and $X, Y, Z \subseteq \Omega$. We sometime use the abbreviate notation $XY$ to represent $X \cup Y$. Let $t[X]$ be the tuple that extracts its component in attributes $X$. The *projection* of $R$ on $X$ is defined as follows:

$$R[X] = \{t[X] | t \in R\}.$$

For instance, truth tables $T_g$ and $T_h$ given in Fig.1(d) and Fig.1(e), respectively, are projections of the truth table $T_f$ in Fig.1(c) on $x_2 x_3 F$ and $x_1 x_4$, respectively.

Similarly, the *conditional projection* of $R$ on $X$ by a $Y$-value $y$ is defined as follows:

$$R_y[X] = \{t[X] | t \in R \text{ and } t[Y] = (y)\}.$$

For example, consider the truth table $T_{wf}$ in Fig.1(g) as a relation $R[x_1 x_4 W x_2 x_3 F]$, the conditional projection $R_{W=1}[x_1 x_4] = \{t_0, t_2, t_9, t_{11}, t_4, t_6, t_{13}, t_{15}\}$ is the set of tuples whose $W$-value are all equal to 1.

The *join* operation is the generalization of Cartesian product to combine two relations with common attributes. Suppose that $X$, $Y$ and $Z$ are disjoint sets of attributes and $S[XY]$ and $T[XZ]$ be two relations. The *join* of $S$ and $T$, denoted by $R[XYZ]$, is defined as follows [21, 25]:

$$
\begin{aligned}
R[XYZ] &= S[XY] \bowtie T[XZ] \\
&= \{(x, y, z) | (x, y) \in S \text{ and } (x, z) \in T\} \\
&= \{S_x[Y] \times T_x[Z] \times (x) | (x) \in S[X] \cap T[X]\}
\end{aligned}
$$

The relation $S[XY] \bowtie T[XZ]$ is formed by taking each tuple $s = (x, y)$ from $S$ and each tuple $t = (x, z)$ from $T$ and combining them on the condition of their $X$-values, a tuple $r = (x, y, z)$ will be formed for $R$ if the components of $s$ and $t$ for attributes $X$ are equal.

### 3.2 Partitions and Bipartite Graphs

The relational approach to the decomposition of a switching function lies on the properties of partitions, which contain groups of equivalence classes of tuples. We portray the interrelationship of two partitions by a bipartite graphs, and show that both FD and MVD can be characterized by bipartite graphs possessing specific topological properties. Our algorithm is a procedure to construct those bipartite graphs of interest to satisfy functional requirements of decomposition.

A partition on a finite nonempty set $S$ is a family $\pi = \{B_1; B_2; \cdots; B_n\}$ of subsets, called blocks, with the following properties [26]:

(1) $B_i \cap B_j = \phi$ if $i \neq j$,
(2) $B_1 \cup B_2 \cup \cdots \cup B_n = S$.

For $a, b \in S$, we write $a\pi b$ to indicate that $a$ and $b$ are in the same block of $\pi$. The set of all partitions on $S$ is a *partial ordered set* (poset) with the following partial ordering:

$$\pi_1 \leq \pi_2$$

if $\forall P_i \in \pi_1$ there exists a $Q_j \in \pi_2$ such that $P_i \subseteq Q_j$.

**Definition 1** *A partition $\pi_X$ on the set of tuples in relation $R[\Omega]$ associated with a subset $X \subseteq \Omega$ of attributes is defined as follows [25]:*

$$\theta : X \to \pi_X = \{t_1 \pi_X t_2 \mid t_1[X] = t_2[X]\}$$

As an example, consider the truth table given in Fig.1(c), the partition $\pi_{x_1 x_4}$ on the tuples of $R[x_1 x_2 x_3 x_4 F]$ contains 4 blocks, and tuples in the same block of $\pi_{x_1 x_4}$ should have the same $x_1 x_4$-value. It is convenient sometime that each block of partition $\pi_Y$ is indexed by its $Y$-value. For instance, blocks of partition $\pi_{x_1 x_4} = \{P_{00}; P_{01}; P_{10}; P_{11}\}$ are indexed by values of $x_1 x_4$, and every tuple in the block $P_{00} = \{t_0, t_2, t_4, t_6\}$ possesses the same value $x_1 x_4 = 00$.

**Definition 2** *The interrelationship between two partitions $\pi_1$ and $\pi_2$ on $S$ can be described by a bipartite graph $G(\pi_1 \times \pi_2, S)$ defined below:*

(1) *blocks of $\pi_1$ and $\pi_2$ are nodes of $G$;*
(2) *for each $s \in S$, there is an edge between $P_i \in \pi_1$ and $Q_j \in \pi_2$ if $s \in P_i \cap Q_j$.*

From the above definition, the following properties of the *greatest lower bound* and the *lowest upper bound* associated with two partitions $\pi_1$ and $\pi_2$ generally hold in the bipartite graph $G(\pi_1 \times \pi_2, S)$:

(1) *Meet* (greatest lower bound): $a\pi_1 \wedge \pi_2 b$ if and only if $a\pi_1 b$ and $a\pi_2 b$. The block $P_i \cap Q_j$ of $\pi_1 \wedge \pi_2$ contains all edges between $P_i$ and $Q_j$;
(2) *Join* (lowest upper bound): $a\pi_1 \vee \pi_2 b$ if and only if there exists a sequence $c_1, c_2, \cdots, c_n \in S$, such that $a\pi_1 c_1 \pi_2 c_2 \pi_1 \cdots \pi_1 c_n \pi_2 b$. Each block of $\pi_1 \vee \pi_2$ contains all edges of a connected component of the bipartite graph $G$.

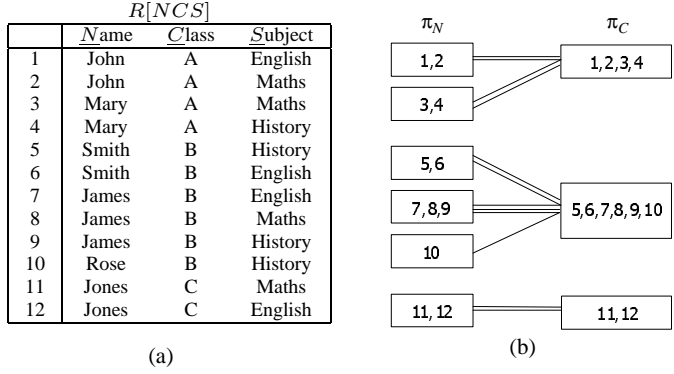

Fig. 2. The functional dependency FD: $N \to C$ on $R[NCS]$.

There are two classes of bipartite graphs of special interest due to their specific topological properties in connection with FDs and MVDs of a relation. The first one defined below is related to FDs.

**Example 2** *A fork representing the interrelationships between two partitions of the course arrangement table $R[NCS]$ is given in Fig.2(b).*

The MVDs are related to the following class of bipartite graph.

**Definition 3** *A bipartite graph $G(\pi_1 \times \pi_2, S)$ is* uniform *if there is one and only one edge between any two blocks $P_i \in \pi_1$ and $Q_j \in \pi_2$ of a connected component of $G$.*

That is, each connected component of a uniform bipartite graph $G$ is a complete bipartite subgraph [27]. As an example, the graph $G(\pi_{WY} \times \pi_{WZF}, R)$ depicted in Fig.1(j) is a uniform bipartite graph, in which each edge represents a tuple (minterm) in $R$. This uniform bipartite graph consists of two complete bipartite subgraphs $K_{2,4}$.

**Definition 4** *Two graphs $G_1 = (V_1, E_1)$ and $G_2 = (V_2, E_2)$ are* isomorphic *if there is a one-to-one and onto mapping $\phi : V_1 \to V_2$, such that for any two vertices $a, b \in V_1$, there is an edge $ab$ in $G_1$ if and only if there is an edge $\phi(a)\phi(b)$ in $G_2$. The mapping $\phi$ is called an* isomorphism.

### 3.3  Functional Dependency

A functional dependency is a constraint between two sets of attributes in a relation $R[\Omega]$. A set of attributes $X$ is said to functionally determine another set of attributes $Y$, written FD: $X \to Y$, if and only if each $X$ value is associated with at most one $Y$ value.

**Definition 5** *The relation $R[\Omega]$ has the functional dependency FD: $X \to Y$,*

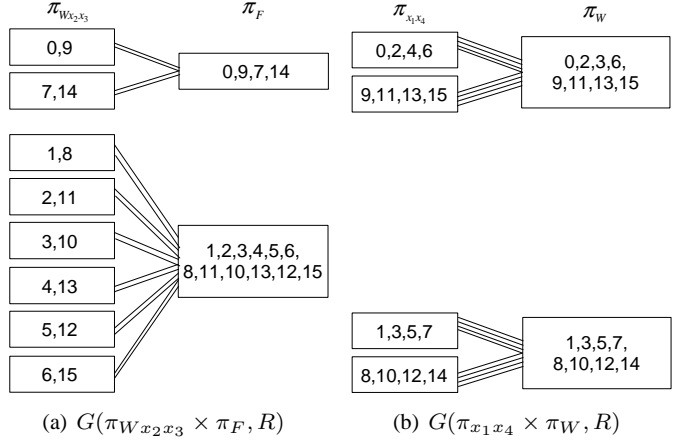

(a) $G(\pi_{Wx_2x_3} \times \pi_F, R)$     (b) $G(\pi_{x_1x_4} \times \pi_W, R)$

Fig. 3. FD: $Wx_2x_3 \to F$ and FD: $x_1x_4 \to W$ of $T_{wh}$ and $T_{wg}$ in Fig.1.

*if $t_1[X] = t_2[X]$ implies $t_1[Y] = t_2[Y]$ $\forall t_1, t_2 \in R$, where $X, Y \subseteq \Omega$.*

**Example 3** *In the course arrangement table $R[NCS]$ shown in Fig.2(a), we observe that the attribute $N$ (Name) determines the attribute $C$ (Class). For example, John is always in class $A$ no matter which subject he took. In other words, the class number is fixed for each student. By definition, there is an FD: $N \to C$ in the table $R[NCS]$. This interrelationship is described by the fork $G(\pi_N \times \pi_C, R)$ depicted in Fig.2(b). In general, the functional dependency FD: $X \to Y$ implies $\pi_X \leq \pi_Y$, or $\pi_{XY} = \pi_X$. That is, the partition $\pi_X$ associated with attributes $X$ is a refinement of the partition $\pi_Y$. It follows immediately that the following lemma should hold.*

**Lemma 1** *The relation $R[\Omega]$ has the functional dependency FD: $X \to Y$, if and only if the bipartite graph $G(\pi_X \times \pi_Y, R)$ is a fork, and blocks of $\pi_Y$ are roots of trees in $G$.*

The concept of functional dependency in a relation is closely related to the function in a truth table. For example, the two forks in Fig.3(a) and (b) represent functional dependencies FD: $Wx_2x_3 \to F$ and FD: $x_1x_4 \to W$, respectively, in the truth table $T_{wf}$ given in Fig.1(g).

### 3.4 Multi-valued Dependency

The multi-valued dependency (MVD) is the necessary and sufficient condition for lossless decomposition of a relation $R[\Omega]$ into two smaller relations. Given two disjoint attribute sets $X$ and $Y$, there is an MVD from $X$ to $Y$ if and only if $R[\Omega]$ is the join of $R[XY]$ and $R[XZ]$, where $Z = \Omega - XY$.

**Definition 6** *A relation $R[\Omega]$ satisfies the MVD: $X \to\to Y$ if, for any $t_1, t_2 \in R$ with $t_1[X] = t_2[X]$, there is a $t_3 \in R$ with $t_3[X] = t_1[X]$, $t_3[Y] = t_1[Y]$, and*

(a) $R[FDP]$

$R[FDP]$

| | $F$light | $D$epart | $P$ilot |
|---|---|---|---|
| 1 | 106 | Mon. | 747 |
| 2 | 106 | Thur. | 747 |
| 3 | 106 | Mon. | 1011 |
| 4 | 106 | Thur. | 1011 |
| 5 | 204 | Wed. | 737 |
| 6 | 204 | Fri. | 737 |
| 7 | 204 | Wed. | 767 |
| 8 | 204 | Fri. | 767 |

(b) Information-lossless decomposition

$R[FDP] =$

$R[FD]$

| | $F$light | $D$epart |
|---|---|---|
| 1 | 106 | Mon. |
| 2 | 106 | Thur. |
| 5 | 204 | Wed. |
| 6 | 204 | Fri. |

⋈

$R[FP]$

| | $F$light | $P$ilot |
|---|---|---|
| 3 | 106 | 747 |
| 4 | 106 | 1011 |
| 7 | 204 | 737 |
| 8 | 204 | 767 |

(c) Redundancy is generated by Cartesian product

$R_{F=106}[DP]$

| | $D$epart | $P$ilot |
|---|---|---|
| 1 | Mon. | 747 |
| 2 | Thur. | 747 |
| 3 | Mon. | 1011 |
| 4 | Thur. | 1011 |

$=$

$R_{F=106}[D]$

| | $D$epart |
|---|---|
| 1,3 | Mon. |
| 2,4 | Thur. |

$\times$

$R_{F=106}[P]$

| | $P$ilot |
|---|---|
| 1,2 | 747 |
| 3,4 | 1011 |

$R_{F=204}[DP]$

| | $D$epart | $P$ilot |
|---|---|---|
| 5 | Wed. | 737 |
| 6 | Fri. | 737 |
| 7 | Wed. | 767 |
| 8 | Fri. | 767 |

$=$

$R_{F=204}[D]$

| | $D$epart |
|---|---|
| 5,7 | Wed. |
| 6,8 | Fri. |

$\times$

$R_{F=204}[P]$

| | $P$ilot |
|---|---|
| 5,6 | 737 |
| 7,8 | 767 |

Fig. 4. Multi-valued dependency MVD: $F \rightarrow\rightarrow D$.

$t_3[Z] = t_2[Z]$. *[21]*

**Example 4** *We elaborate the above definition on multi-valued dependency by the MVD: $F \rightarrow\rightarrow D$ in the airline information table given in Fig.4(a). Specifically, we will demonstrate the equivalence between the MVD: $F \rightarrow\rightarrow D$ and the following information-lossless decomposition*

$$R[FDP] = R[FD] \bowtie R[FP]. \tag{1}$$

*The MVD: $F \rightarrow\rightarrow D$ in relation $R[FDP]$ guarantees that following equalities hold*

$$R_{F=106}[PD] = R_{F=106}[P] \times R_{F=106}[D],$$

*and*

$$R_{F=204}[PD] = R_{F=204}[P] \times R_{F=204}[D]. \tag{2}$$

*A similar expression of the original relation $R[FDP]$ in terms of conditional projections is given by*

$$R[FDP] = (106) \times R_{F=106}[PD] \cup (204) \times R_{F=204}[PD]. \tag{3}$$

*Substitute (2) into (3), we have*

$$\begin{aligned} R[FDP] = &(106) \times R_{F=106}[P] \times R_{F=106}[D] \\ &\cup (204) \times R_{F=204}[P] \times R_{F=204}[D]. \end{aligned} \tag{4}$$

*Recall that $R[FD]$ and $R[FP]$ are projections of relation $R[FDP]$ on attribute sets $FD$ and $FP$, respectively, and they can be expressed by the union of conditional projections as follows*

$$R[FD] = (106) \times R_{F=106}[D] \cup (204) \times R_{F=204}[D],$$

*and*

$$R[FP] = (106) \times R_{F=106}[P] \cup (204) \times R_{F=204}[P]. \tag{5}$$

*According to the definition of join, we know from (5) that*

$$\begin{aligned} R[FP] \bowtie R[FD] = &(106) \times R_{F=106}[P] \times R_{F=106}[D] \\ &\cup (204) \times R_{F=204}[P] \times R_{F=204}[D]. \end{aligned} \tag{6}$$

*The equality (1) can now be established by comparing (4) and (6), in which the Cartesian product $(F = f) \times R_F[P] \times R_F[D]$ for each given value $F = f$ can be explicitly displayed by a complete bipartite graph. For example, the two connected components of $G(\pi_{FD} \times \pi_{FP}, R)$, depicted in Fig.5, correspond to Cartesian products of $R_{F=106}[PD]$ and $R_{F=204}[PD]$, respectively.*

*The relationship between MVD and uniform bipartite graph observed from the above example on airline information table $R[FDP]$ is summarized in the following lemma.*

**Lemma 2** *A relation $R[XYZ]$ has an multi-valued dependency MVD: $X \rightarrow\rightarrow Y$, if and only if $G(\pi_{XY} \times \pi_{XZ}, R)$ is a uniform bipartite graph, in which each connected component of $G$ corresponds to an $X$ value and vise versa.*

**Proof 1** *From the definition of multi-valued dependency, the MVD: $X \rightarrow\rightarrow Y$ in relation $R[XYZ]$ is equivalent to the following identity*

$$R_x[YZ] = R_x[Y] \times R_x[Z], \tag{7}$$

*for each value $X = x$. Perform union of (7) over all $X$ values, it follows immediately that*

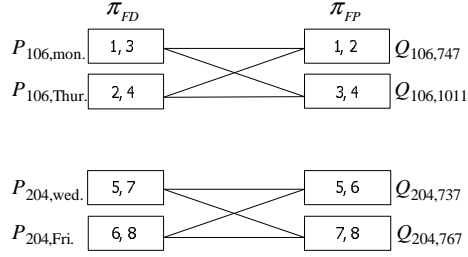

Fig. 5. The uniform bipartite graph $G(\pi_{FD} \times \pi_{FP}, R)$ of the MVD: $F \rightarrow\rightarrow D$ in $R[FDP]$ shown in the Fig.4.

$$R[XYZ] = \{(x) \times R_x[Y] \times R_x[Z] | (x) \in R[X]\} \qquad (8)$$
$$= R[XY] \bowtie R[XZ]. \qquad (9)$$

*Each Cartesian product appeared in (8) is a complete bipartite graph for any given value $X = x$, it is clear that the above expression represents the uniform bipartite graph $G(\pi_{XY} \times \pi_{XZ}, R)$, and the assertion of the lemma is established.*

Return to the running example of Ashenhurst-Curtis decomposition discussed in section 2, the fact that $G(\pi_{WY} \times \pi_{WZF}, R)$ given in Fig.1(j) is a uniform bipartite graph implies that the truth table $T_{wf}$ has MVD: $W \rightarrow\rightarrow x_1 x_4$ which ensures the lossless decomposition of $T_{wf}$ into $T_{wg}$ and $T_{wh}$ according to the above lemma.

## 4    Disjoint Decomposition

Recall that a logic circuit is a function $f$ from an input set $X = \{x_1, x_2, \cdots, x_n\}$ to an output $F$. Given the bound set $Y$ and the free set $Z$ such that $Y \cup Z = X$ and $Y \cap Z = \phi$, the *disjoint decomposition* of $f$ into two functions $W = g(Y)$ and $F = h(W, Z)$ aims at finding the partition $\pi_W$ of a set of bridge variables $W$ such that the truth table $R[WXF] = R[WYZF]$ satisfies the following:

(1) FD: $Y \rightarrow W$.
(2) FD: $WZ \rightarrow F$.
(3) MVD: $W \rightarrow\rightarrow Y$, or equivalently, $R[WYZF] = R[WY] \bowtie R[WZF]$.

If $\pi_W$ has $k$ blocks, then it requires $\lceil \log_2 k \rceil$ bits to encode blocks of $\pi_W$, in which case $|W| = \lceil \log_2 k \rceil$. It is clear that a decomposition is *nontrivial* only when $\lceil \log_2 k \rceil < |Y|$. The procedure of computing the partition $\pi_W$ is described in the sequel, and we also prove that the proposed decomposition algorithm is information-lossless.

The proposed disjoint functional decomposition algorithm (FDA) is based on a decomposition chart, denoted by $M_{ZY}$, which is constructed from the truth table $R[YZF]$ as follows:

(1) The columns, called *bound domain*, of $M_{ZY}$ are named by blocks $P_y \in \pi_Y$ ($y$ is the $Y$-value of block $P_y$), the partition induced by $Y$.
(2) The rows, called *free domain*, of chart $M_{ZY}$ are named by blocks $Q_z$ ($z$ is the $Z$-value of block $P_z$), the partition induced by $Z$.
(3) Since $\pi_Y \wedge \pi_Z = 0$, the intersection of $Q_z$ and $P_y$ contains a single tuple $t$ of the truth table $R[YZF]$. For $t \in P_y \cap Q_z$, the entry $(Q_z, P_y)$ is the $F$-value of $t$ in the truth table.

**Definition 7** *In the chart $M_{ZY}$, we say two columns $P_i$ and $P_j$ are* equivalent, *if two tuples associated with $(Q_z, P_i)$ and $(Q_z, P_j)$ have the same $F$-value for all $z$ in the free domain. A set of columns is called an* equivalent column set, *if columns in the set are all equivalent.*

Based on chart $M_{ZY}$, the truth table $R[YZF]$ is decomposed by the following procedure:

**Disjoint Decomposition Algorithm (FDA-$\alpha$)**

S1. (*Initialization*) Establish the chart $M_{ZY}$ from the truth table $R[YZF]$.
S2. Exhaustively merge equivalent columns to form a single column, until there are no equivalent columns.
S3. The columns of the final chart, denoted $M_{ZW}$, are blocks of $\pi_W$.

Notice that the merge of two equivalent columns $P_i$ and $P_k$ to form column $P_{i \vee j}$ in S2 ensures that all tuples in the entry $Q_z \cap P_{i \vee j} = (Q_z \cap P_i) \cup (Q_z \cap P_j)$ have the same $F$-value. That is, S2 eliminates all duplicated columns, redundancies, of the chart $M_{ZY}$, and the final chart $M_{ZW}$ contains the same information of the truth table $R[YZF]$ as the original chart.

The above procedure will be elaborated by the running example of Ashenhurst-Curtis decomposition given in Section 2. There are two equivalent column sets $\{P_{00}, P_{11}\}$ and $\{P_{01}, P_{10}\}$ in the decomposition chart $M_{ZY}$ constructed from the truth table given in Fig.6(a). The final chart $M_{ZW}$, displayed in Fig.6(b), is obtained by merging equivalent columns in $M_{ZY}$. Blocks of partition $\pi_W$ are columns $P_{00 \vee 11}$ and $P_{01 \vee 10}$ of the final chart $M_{ZW}$, i.e., $\pi_W = \{P_{00 \vee 11}, P_{01 \vee 10}\}$. Assigning $W$ values "$W = 1$" and "$W = 0$" to tuples in $P_{00 \vee 11}$ and $P_{01 \vee 10}$, respectively, we finally obtain the truth table $T_{wf}$, displayed in Fig.1(g), with the bridge variable $W$ being introduced. The projection of $T_{wf}$ on the set of

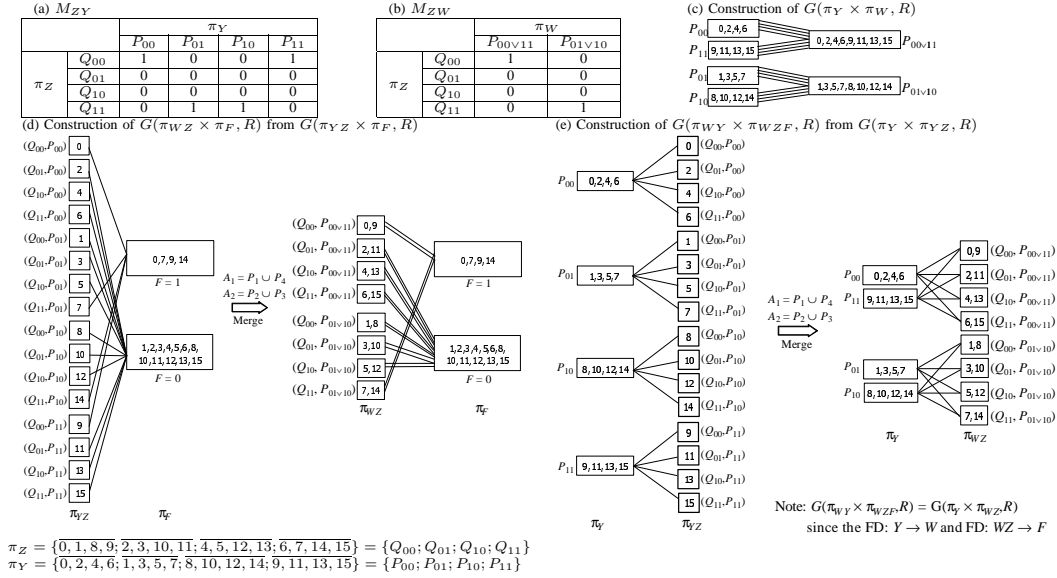

Fig. 6. Decomposition chart $M_{ZY}$ and final chart $M_{ZW}$ of example 1.

variables $\{x_2, x_3, W, F\}$, shown in Fig.1(h), is the truth table of the function

$$F = h(W, Z) = h(W, x_2, x_3) = W\overline{x}_2\overline{x}_3 + \overline{W}x_2x_3.$$

Similarly, the projection of $T_{wf}$ on the set of variables $\{x_1, x_4, W\}$, shown in Fig.1(i), is the truth table of the function

$$W = g(Y) = g(x_1, x_4) = \overline{x}_1\overline{x}_4 + x_1x_4.$$

This decomposition is nontrivial, because blocks of $\pi_W$ can be encoded by 1 bit, less than $|Y| = 2$.

### 4.2 Information-lossless Property of Disjoint Decomposition

We will prove that the proposed FDA-$\alpha$ is essentially a procedure to construct the uniform bipartite graph associated with the MVD: $W \rightarrow\rightarrow Y$, which ensures the information-lossless decomposition of the underlying logic circuit.

**Theorem 1** *The set of bridge variables $W$ obtained by FDA-$\alpha$ satisfies the following properties:*

*(1) FD: $Y \rightarrow W$.*
*(2) FD: $WZ \rightarrow F$.*
*(3) MVD: $W \rightarrow\rightarrow Y$.*
*(4) For any $W^*$ that satisfies the properties (1)–(3), we have $\pi_{W^*} \leq \pi_W$.*

**Proof 2** *Without loss of generality, we use the Ashenhurst-Curtis decomposition example given in Section 2 to prove this theorem.*

*(1) FD: $Y \rightarrow W$. Because each block of $\pi_W$ is the union of several blocks of $\pi_Y$, it is obvious that $\pi_Y \leq \pi_W$, which is equivalent to the FD: $Y \rightarrow W$, as illustrated by the fork $G(\pi_Y \times \pi_W, R)$ in Fig.6(c).*

*(2) FD: $WZ \rightarrow F$. The fork $G(\pi_{YZ} \times \pi_F, R)$, shown in the left end of Fig.6(d), represents the functional dependency FD: $YZ \rightarrow F$ in the truth table $R[YZF]$. In S2 of FDA-$\alpha$, blocks (equivalent columns) of $\pi_Y$ are merged to form blocks of $\pi_W$. As a result, the fork $G(\pi_{YZ} \times \pi_F, R)$ becomes the fork $G(\pi_{WZ} \times \pi_F, R)$ shown on the right end of Fig.6(d). According to lemma 1, this fork $G(\pi_{WZ} \times \pi_F, R)$ represents the FD: $WZ \rightarrow F$. For instance, consider the two tuples $t_0, t_9 \in (Q_{00}, P_{00 \vee 11}) = (Q_{00}, P_{00}) \cup (Q_{00}, P_{11})$, we have*

$$t_0 \pi_{WZ} t_9.$$

*Given that the two columns $P_{00}$ and $P_{11}$ are equivalent, it follows that*

$$t_0 \pi_F t_9,$$

*specifically, $t_0[F] = t_9[F] = 1$. Note that this fork $G(\pi_{WZ} \times \pi_F, R)$ also appears in Fig.3(a) before.*

*(3) MVD: $W \rightarrow\rightarrow Y$. The above functional dependencies FD: $Y \rightarrow W$ and FD: $WZ \rightarrow F$ imply $\pi_{WY} = \pi_Y$ and $\pi_{WZF} = \pi_{WZ}$, respectively. Therefore, it is obvious that*

$$G(\pi_{WY} \times \pi_{WZF}, R) = G(\pi_Y \times \pi_{WZ}, R).$$

*According to lemma 2, the MVD: $W \rightarrow\rightarrow Y$ holds in $R[WYZF]$ if the bipartite graph $G(\pi_Y \times \pi_{WZ}, R)$ is uniform. Since $\pi_{YZ} \leq \pi_Y$, the bipartite graph $G(\pi_Y \times \pi_{YZ}, R)$ is a fork consisting of 4 trees, as depicted in Fig.6(e). The root of each tree corresponds to a column of the chart $M_{ZY}$, and each leaf is a block of $\pi_{YZ}$ representing an entry in $M_{ZY}$. Thus, "equivalent" trees (columns) have same number of leaves. In S2 of FDA-$\alpha$, equivalent columns are merged to form blocks of $\pi_W$. It follows from this merge operation that corresponding leaves of equivalent trees (columns) in $G(\pi_Y \times \pi_{YZ}, R)$ are glued together to form the bipartite graph $G(\pi_Y \times \pi_{WZ}, R)$, as shown in Fig.6(e), which is uniform simply because equivalent trees are* isomorphic, *i.e., topologically identical.*

*(4) For any $W^*$ that satisfies the properties (1)–(3), we have $\pi_{W^*} \leq \pi_W$.*

*Suppose the set of bridge variables $W^*$ satisfies properties (1)–(3), then the FD: $Y \rightarrow W^*$ of (1) implies $\pi_Y \leq \pi_{W^*}$. That is, each block of partition $\pi_{W^*}$ is* a set union (merge) of several blocks of the partition $\pi_Y$. We claim that each of these blocks of $\pi_{W^*}$ must be the merge of equivalent columns of the chart $M_{ZY}$. This assertion will be proved by a contradiction below.

Assume that the block $P_{i\vee j}$ of $\pi_{W^*}$ is the merge of two columns $P_i$ and $P_j$ which are not equivalent, then there exists a row $Q_z$ such that two tuples in $(Q_z, P_i)$ and $(Q_z, P_j)$ possess different $F$-values in the chart $M_{ZY}$. It follows that these two tuples appearing in $(Q_z, P_{i\vee j}) = (Q_z, P_i) \cup (Q_z, P_j)$ will connect to two different roots in the bipartite graph $G(\pi_{W^*Z} \times \pi_F, R)$, which can no longer be a fork, and violate FD: $W^*Z \to F$, the property (2) of $W^*$. Thus, the assumption will lead to a contradiction because the property (3) of $W^*$ can not be held either. An example of this $F$-value conflict is shown in Fig.6(d), in which the bipartite graph $G(\pi_{WZ} \times \pi_F)$ would not be a fork if $P_{00}$ and $P_{01}$ in $G(\pi_{YZ} \times \pi_F)$ are merged together, because the two tuples $t_6, t_7 \in (Q_{11}, P_{00}) \cup (Q_{11}, P_{01})$, possessing different $F$-values, will be connected to two different roots on the right end.

Because the set of bridge variables $W$ is obtained by the exhaustive merge of all equivalent columns in $S2$ of FDA-$\alpha$, there should be no more equivalent columns left in the final chart $M_{ZW}$. Thus, for any $W^*$ that satisfies the properties (1)–(3), we have $\pi_{W^*} \leq \pi_W$.

## 5   Multiple Decomposition

It is common in practical application that a complex logic circuit may have to be decomposed into several smaller circuits, called *multiple decomposition*. Given a set of pairwise mutually disjoint subsets $Y_1 \cup Y_2 \cdots \cup Y_K \cup Z = X$, the multiple decomposition of $f$ into a series of functions $W_1 = g_1(Y_1), W_2 = g_2(Y_2), \cdots, W_K = g_K(Y_K)$, and $F = h(W_1, W_2, \cdots, W_K, Z)$ is to introduce $K$ sets of bridge variables $W_1, W_2, \cdots, W_K$ such that the truth table $R[W_1 W_2 \cdots W_K X F] = R[W_1 W_2 \cdots W_K Y_1 Y_2 \cdots Y_K Z F]$ satisfies the following properties:

(1) FDs: $Y_1 \to W_1, Y_2 \to W_2, \cdots, Y_K \to W_K$.
(2) FD: $W_1 W_2 \cdots W_K Z \to F$.
(3) MVD: $W_1 \to\to Y_1, W_2 \to\to Y_2, \cdots, W_K \to\to Y_K$.

**Example 5** *Similar to the FDA-$\alpha$ of simple disjoint decomposition, the procedure of multiple decomposition is illustrated by the example shown in Fig.7. Let $Y_1 = \{x_1, x_4, x_5\}$ and $Y_2 = \{x_2, x_3\}$ be free sets in turn, and establish charts $M_{Y_2 Y_1}$ and $M_{Y_1 Y_2}$ according to the given truth table. Apply the FDA-$\alpha$ to both charts, we obtain bridge variables $W_1$ and $W_2$, respectively. Based on theorem 1, it is easy to show that the multiple decomposition is also lossless.*

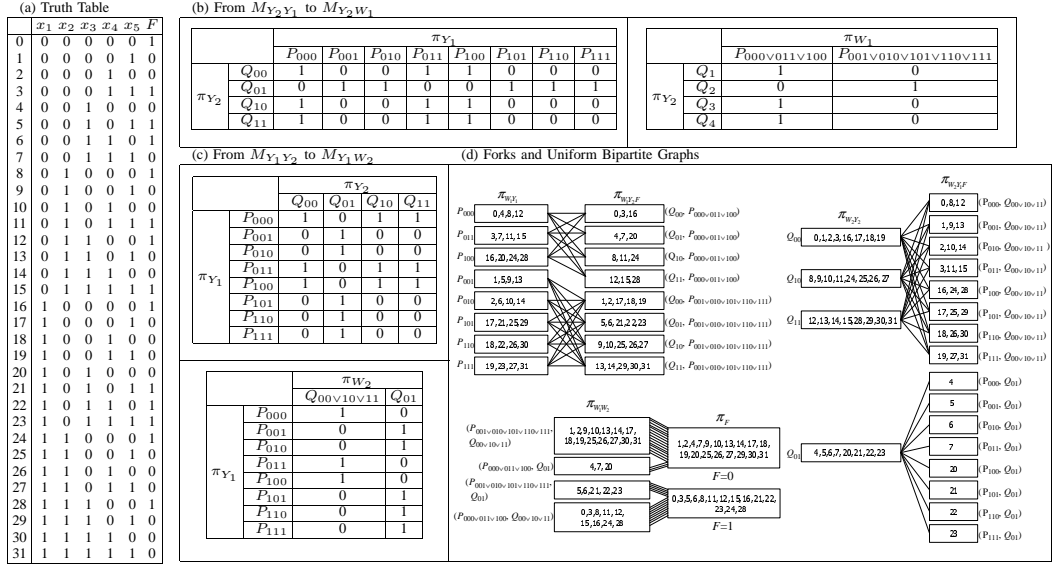

Fig. 7. Multiple decomposition with $Y_1 = \{x_1, x_4, x_5\}$ and $Y_2 = \{x_2, x_3\}$.

*Furthermore, the uniform bipartite graphs $G(\pi_{W_1Y_1} \times \pi_{W_1Y_2F}, R)$ and $G(\pi_{W_2Y_2} \times \pi_{W_2Y_1F}, R)$ exhibit the MVD: $W_1 \to\to Y_1$ and MVD: $W_2 \to\to Y_2$, respectively, while the fork $G(\pi_{W_1W_2} \times \pi_F, R)$ represents the FD: $W_1W_2 \to F$. Assigning "$W_1 = 1$" and "$W_1 = 0$" to tuples in $P_{000\vee011\vee100}$ and $P_{001\vee010\vee101\vee110\vee111}$, blocks of partition $\pi_{W_1}$, yields*

$$W_1 = g_1(x_1, x_4, x_5) = \overline{x}_1 x_4 x_5 + \overline{x}_4 \overline{x}_5.$$

*Similarly, assigning "$W_2 = 1$" and "$W_2 = 0$" to tuples in $Q_{00\vee10\vee11}$ and $Q_{01}$, blocks of partition $\pi_{W_2}$, gives*

$$W_2 = g_2(x_2, x_3) = x_2 + \overline{x}_3.$$

*The truth table of the function $F = h(W_1, W_2)$ is essentially a chart $M_{W_1W_2}$ that can be obtained by either merge equivalent rows of chart $M_{Y_2W_1}$, or merge equivalent rows of chart $M_{Y_1W_2}$. With respect to the previous assignments of $W_1$ and $W_2$ values, the function $F = h(W_1, W_2)$ is given below*

$$F = h(W_1, W_2) = \overline{W}_1 \overline{W}_2 + W_1 W_2.$$

*The schematic diagram of the decomposed function $F = f(X)$, consisting of $F = h(W_1, W_2)$, $W_1 = g_1(Y_1)$ and $W_2 = g_2(Y_2)$, is shown in Fig.8.*

The multiple decomposition procedure is listed as follows:
**Multiple Decomposition Algorithm (FDA-$\beta$)**

S1. (*Initialization*) Set $k = 1$, and let $Z_k = X - Y_k$.

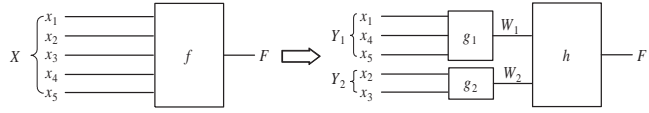

Fig. 8. Decompose $F = f(x_1, x_2, x_3, x_4, x_5)$ into $F = h(g_1, g_2)$, $W_1 = g_1(x_1, x_4, x_5)$, and $W_2 = g_2(x_2, x_3)$.

S2. Establish chart $M_{Z_k Y_k}$ from the given truth table, where $Y_k$ is the bound set and $Z_k$ is the free set.

S3. Exhaustively merge equivalent columns in the chart $M_{Z_k Y_k}$ to determine partition $\pi_{W_k}$.

S4. $k = k + 1$. If $k \leq K$, go to S2; otherwise, terminate the algorithm.

# 6    Non-disjoint Decomposition

An extension of FDA-$\alpha$ to the *non-disjoint decomposition* is described in this section. The non-disjoint decomposition is different from the disjoint decomposition in that the intersection of the bound set $Y$ and the free set $Z$ is nonempty, i.e., $Y \cap Z = C$ where $C \neq \phi$. The procedure of the non-disjoint decomposition of a function $f$ is illustrated by the following example.

## 6.1    Non-disjoint decomposition Algorithm

**Example 6** *Consider the truth table of a logic circuit given in Fig.9(a), and let the bound set be $Y = \{x_2, x_4, x_5\}$ and the free set be $Z = \{x_1, x_2, x_3\}$. Condition on $Y \cap Z = \{x_2\}$, the conditional projections of the the truth table $R[YZF]$ are defined as follows*

$$R_{x_2=0}[x_1 x_3 x_4 x_5 F] = R_{x_2=0}[(Y - x_2)(Z - x_2)F]$$

*and*

$$R_{x_2=1}[x_1 x_3 x_4 x_5 F] = R_{x_2=1}[(Y - x_2)(Z - x_2)F]$$

*with truth tables given in Fig.9(b) and (c), respectively. The non-disjoint decomposition of $R[YZF]$ is then reduced to disjoint decompositions of $R_{x_2=0}[(Y - x_2)(Z - x_2)F]$ and $R_{x_2=1}[(Y - x_2)(Z - x_2)F]$, that can be conducted by the FDA-$\alpha$ independently.*

*The difference between the decomposition chart $M_{ZY}$ and the counterpart in the disjoint decomposition is that the chart $M_{ZY}$ of non-disjoint decomposition* contains null entries, e.g., the entry $P_{000} \cap Q_{010} = \phi$ in Fig.9(d) is empty. In chart $M_{ZY}$, the crosses of rows and columns belonging to different blocks of $\pi_{x_2}$ are all null entries. Gathering the rows and the columns in the same block of $\pi_{x_2}$ together, the chart $M_{ZY}$ in Fig.9(d) can be organized in diagonal form, which consists of $|\pi_{x_2}| = 2$ sub-charts in the diagonal. They are corresponding to the truth tables given in Fig.9(b) and (c), respectively. The S2 merge operation of FDA-$\alpha$ can now be operated on equivalent columns in each of these sub-charts independently to obtain an intermediate chart $M_{ZV}$ shown in Fig.9(e).

Fig. 9. Non-disjoint decomposition with $Y = \{x_2, x_4, x_5\}$ and $Z = \{x_1, x_2, x_3\}$.

The sub-charts in the the diagonal of $M_{ZV}$ are orthogonal, merge columns of different sub-charts would not cause any $F$-value conflicts. It is also clear that columns of the same sub-chart in $M_{ZV}$ can no longer be merged, and the number of blocks in partition $\pi_W$ is determined by the maximal number columns over these sub-charts. However, there are many alternative combinations, depending on the choice of the column in each sub-chart, to merge those orthogonal sub-charts. For example, the following final charts are resulting from two legitimate combinations of columns in the chart $M_{ZV}$ in Fig.9(e):
Case 1: The chart $M_{ZW}$ in Fig.9(f)

$$\begin{aligned}
\pi_W &= \{P_{000\vee010\vee011\vee110\vee111}; P_{001\vee100\vee101}\} \\
&= \{P_{000\vee010\vee011} \cup P_{110\vee111}; P_{001} \cup P_{100\vee101}\} \\
&= \{P_{000} \cup P_{010} \cup P_{011} \cup P_{110} \cup P_{111}; P_{001} \cup P_{100} \cup P_{101}\}.
\end{aligned}$$

Case 2: The chart $M_{ZW'}$ in Fig.9(f)

$$\pi_{W'} = \{P_{000 \vee 010 \vee 011 \vee 100 \vee 101}; P_{001 \vee 110 \vee 111}\}$$
$$= \{P_{000 \vee 010 \vee 011} \cup P_{100 \vee 101}; P_{001} \cup P_{110 \vee 111}\}$$
$$= \{P_{000} \cup P_{010} \cup P_{011} \cup P_{100} \cup P_{101}; P_{001} \cup P_{110} \cup P_{111}\}.$$

*The following subfunctions can be obtained from the partition $\pi_W$ of case 1 above, with $W$-values "$W = 0$" and "$W = 1$" being assigned to tuples in $P_{000 \vee 010 \vee 011 \vee 110 \vee 111}$ and $P_{001 \vee 100 \vee 101}$, respectively,*

$$W = g(x_2, x_4, x_5) = \overline{x}_2 \overline{x}_5 + x_4,$$

*and*

$$F = h(W, x_1, x_2, x_3) = W \overline{x}_1 \overline{x}_3 + \overline{W} x_1 + x_1 x_3.$$

The procedure of non-disjoint decomposition of a switch function $f$, called FDA-$\gamma$, is summarized as follows.

**Non-disjoint Decomposition Algorithm (FDA-$\gamma$)**

S1. (*Initialization*) Establish the decomposition chart $M_{ZY}$ in diagonal form according to partition $\pi_C$.

S2. Exhaustively merge equivalent columns of each sub-chart in the diagonal of $M_{ZY}$ to obtain the intermediate chart $M_{ZV}$.

S3. Exhaustively merge columns chosen from different sub-charts, until further merge is impossible, to obtain the final chart $M_{ZW}$.

S4. The number of blocks in $\pi_W$ should be equal to $\lceil \log_2 \lambda \rceil$, where $\lambda = \max_{k \le |\pi_C|}(N_k)$ and $N_k$ is the number of columns in the $k$-th sub-chart $(k \le |\pi_C|)$ of $M_{ZV}$.

*6.2 Information-lossless Property of Non-disjoint Decomposition*

The information-lossless property of non-disjoint decomposition algorithm (FDA-$\gamma$) is detailed in the following corollary.

**Corollary 1** *The set of bridge variables $W$ obtained by FDA-$\gamma$ satisfies the following properties:*

*(1) FD: $Y \rightarrow W$.*

*(2) FD: $WZ \rightarrow F$.*

*(3) MVD: $WC \rightarrow\rightarrow (Y - C)$, or equivalently, $R[WYZF] = R[WY] \bowtie R[WZF]$.*

*(4) For any $W^*$ that satisfies the properties (1)–(3), we have $|W| \le |W^*|$.*

**Proof 3** *Again, without loss of generality, the proof of this corollary is established by the elaboration of Example 6.*

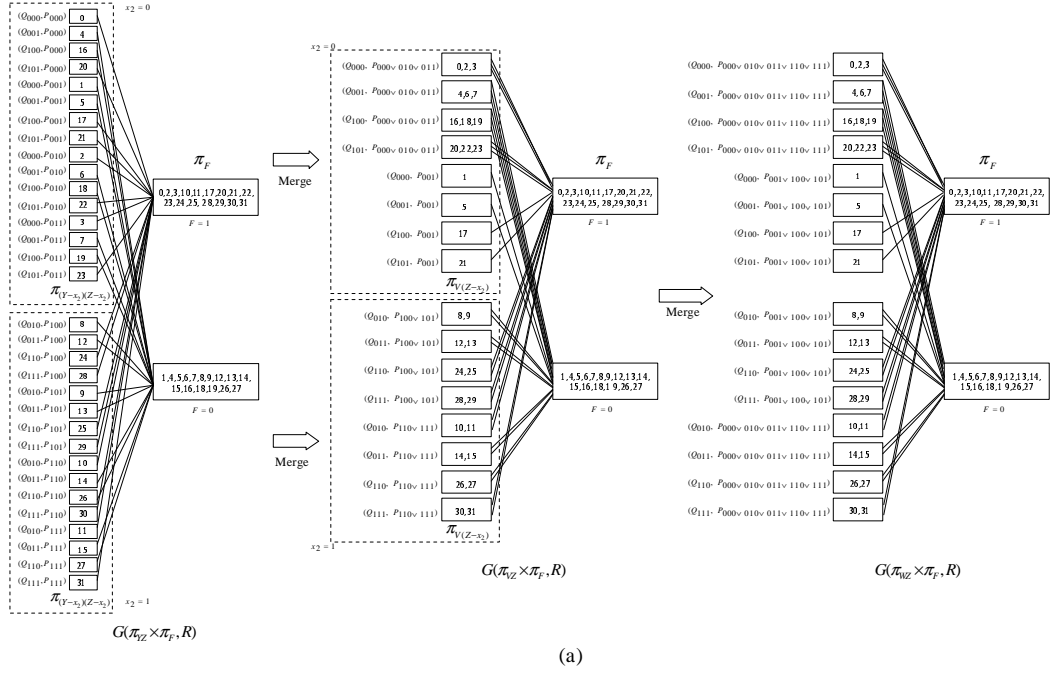

(a)

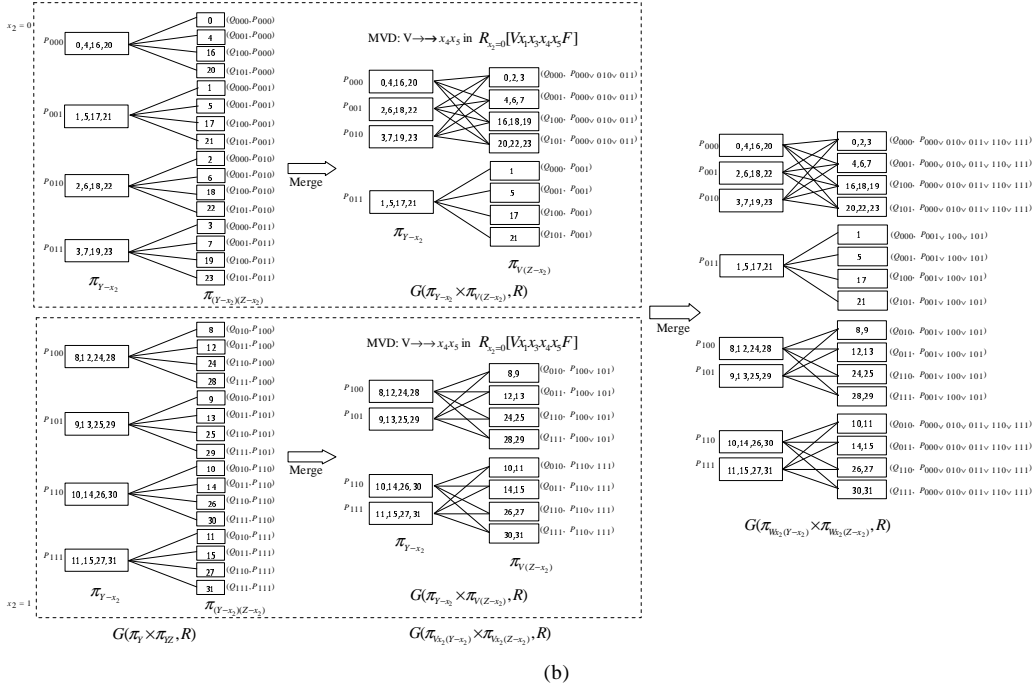

(b)

Fig. 10. Procedure of constructing forks and uniform bipartite graphs for the example shown in Fig.9.

(1) FD: $Y \to W$. From part (1) of theorem 1, the FD: $(Y - x_2) \to V$ holds in both truth tables $R_{x_2=0}[V(Y-x_2)(Z-x_2)F]$ and $R_{x_2=1}[V(Y-x_2)(Z-x_2)F]$. Since the blocks of partition $\pi_W$ are set unions of the blocks of partition $\pi_V$. It is obvious that the FD: $Y \to W$ is valid in $R[WYZF]$ for each value of $x_2$.

(2) FD: $WZ \to F$. Part (2) of theorem 1 ensures that both truth tables

$R_{x_2=0}[V(Y-x_2)(Z-x_2)F]$ and $R_{x_2=1}[V(Y-x_2)(Z-x_2)F]$ have the FD: $V(Z-x_2) \to F$. Thus, the FD: $VZ \to F$ should hold in $R[VYZF]$ for each value of $x_2$, the fork shown in the middle of Fig.10(a) represents this functional dependency.

The FD: $WZ \to F$ holds in $R[WYZF]$ simply because the two forks $G(\pi_{VZ} \times \pi_F, R)$ and $G(\pi_{WZ} \times \pi_F, R)$ are isomorphic. Notice that the fork $G(\pi_{WZ} \times \pi_F, R)$ on the right end of Fig.10(a) is identical to the middle fork representing FD: $VZ \to F$, except labels of their leaves. In fact, there is a one-to-one correspondence between sets of leaves of these two forks, e.g., one of such correspondences is given by the following identity:

$$
\begin{aligned}
& (Q_{000}, P_{000\vee010\vee011\vee110\vee111}) \\
= & (Q_{000}, P_{000\vee010\vee011}) \cup (Q_{000}, P_{110\vee111}) \\
= & (Q_{000}, P_{000\vee010\vee011}),
\end{aligned}
\tag{10}
$$

in which the block $P_{000\vee010\vee011\vee110\vee111}$ of partition $\pi_W$ is the merge of columns $P_{000\vee010\vee011}$ and $P_{110\vee111}$ belonging to different sub-charts of $M_{ZV}$. Since sub-charts are orthogonal to each other, the crosses of row $Q_{000}$ with those columns merged to $P_{000\vee010\vee011}$ are all empty except one such as the entry $(Q_{000}, P_{000\vee010\vee011})$ in (10). The identity (10), held for every entry in $M_{ZW}$, genuinely constitutes the one-to-one mapping, the isomorphism, between these two forks.

(3) MVD: $Wx_2 \to\to (Y-x_2)$. The uniform bipartite graphs within the dotted rectangles in Fig.10(b) demonstrate that the MVD: $V \to\to (Y-x_2)$ holds in both $R_{x_2=0}[V(Y-x_2)(Z-x_2)F]$ and $R_{x_2=1}[V(Y-x_2)(Z-x_2)F]$ according to theorem 1, and can be expressed explicitly as follows

$$
\begin{aligned}
& R_{x_2=0}[V(Y-x_2)(Z-x_2)F] = \\
& R_{x_2=0}[V(Y-x_2)] \bowtie R_{x_2=0}[V(Z-x_2)F],
\end{aligned}
\tag{11}
$$

and

$$
\begin{aligned}
& R_{x_2=1}[V(Y-x_2)(Z-x_2)F] = \\
& R_{x_2=1}[V(Y-x_2)] \bowtie R_{x_2=1}[V(Z-x_2)F].
\end{aligned}
\tag{12}
$$

Recall that the truth table $R[VYZF]$ can be written as the union of conditional projections as follows

$$
\begin{aligned}
R[VYZF] = & R[Vx_2(Y-x_2)(Z-x_2)F] \\
= & (x_2=0) \times R_{x_2=0}[V(Y-x_2)(Z-x_2)F] \\
& \cup (x_2=1) \times R_{x_2=1}[V(Y-x_2)(Z-x_2)F].
\end{aligned}
\tag{13}
$$

*Substituting (11) and (12) into (13) yields the following information-lossless decomposition equality:*

$$R[VYZF] = R[Vx_2(Y - x_2)(Z - x_2)F]$$
$$= R[Vx_2(Y - x_2)] \bowtie R[Vx_2(Z - x_2)F]$$
$$= R[VY] \bowtie R[VZF],$$

*which implies that the truth table $R[VYZF]$ has the MVD: $Vx_2 \rightarrow\rightarrow (Y - x_2)$. Furthermore, the MVD: $Vx_2 \rightarrow\rightarrow (Y - x_2)$ is portrayed by the uniform bipartite graph*

$$G(\pi_{VY} \times \pi_{VZF}, R) = G(\pi_{Vx_2(Y-x_2)} \times \pi_{Vx_2(Z-x_2)F}, R)$$

*shown in Fig.10(b), which is the union of two uniform bipartite subgraphs representing the information-lossless decompositions specified in (11) and (12).*

*From the same one-to-one correspondence defined by (10) for every entry in $M_{ZW}$, the two uniform bipartite graphs $G(\pi_{Vx_2(Y-x_2)} \times \pi_{Vx_2(Z-x_2)F}, R)$ and $G(\pi_{Wx_2(Y-x_2)} \times \pi_{Wx_2(Z-x_2)F}, R)$ depicted in Fig.10(b) are isomorphic. It follows that the MVD: $Wx_2 \rightarrow\rightarrow (Y - x_2)$ is valid in the truth table $R[WYZF]$.*

*(4) For any $W^*$ that satisfies the properties (1)–(3), we have $|W| \leq |W^*|$. The relation $\pi_{W^*} \leq \pi_W$ presented in part (4) of theorem 1 does not hold in FDA-$\gamma$, because $\pi_W$ and $\pi_{W^*}$ may be incomparable as there are various combinations to merge columns in S3. Nevertheless, the number of blocks in $\pi_W$ has been minimized by the exhaustive merge operations in S2 and S3 of FDA-$\gamma$. Thus, $|W| \leq |W^*|$.*

## 7 Incompletely Specified Function

Often, some $F$-values of a function $f(X)$ are unspecified, that is, for an $X$-value, the associated $F$-value of the function $f$ can be arbitrarily assigned by a value from its range. The $f : X \rightarrow F$ is then referred to as *an incompletely specified function.*

**Example 7** *The truth table shown in Fig.11(a) is an example of incompletely specified function, in which a tuple with its $F$-value being assigned by "-" is an unspecified tuple, called "don't care", whose $F$-value can be arbitrarily set to either "0" or "1". In this section, we describe the extension of FDA-$\alpha$ to the decomposition of incompletely specified functions. The procedure is elaborated by the decomposition of the truth table in Fig.11(a) for given bound set $Y = \{x_1, x_2, x_4\}$ and free set $Z = \{x_3, x_5\}$.*

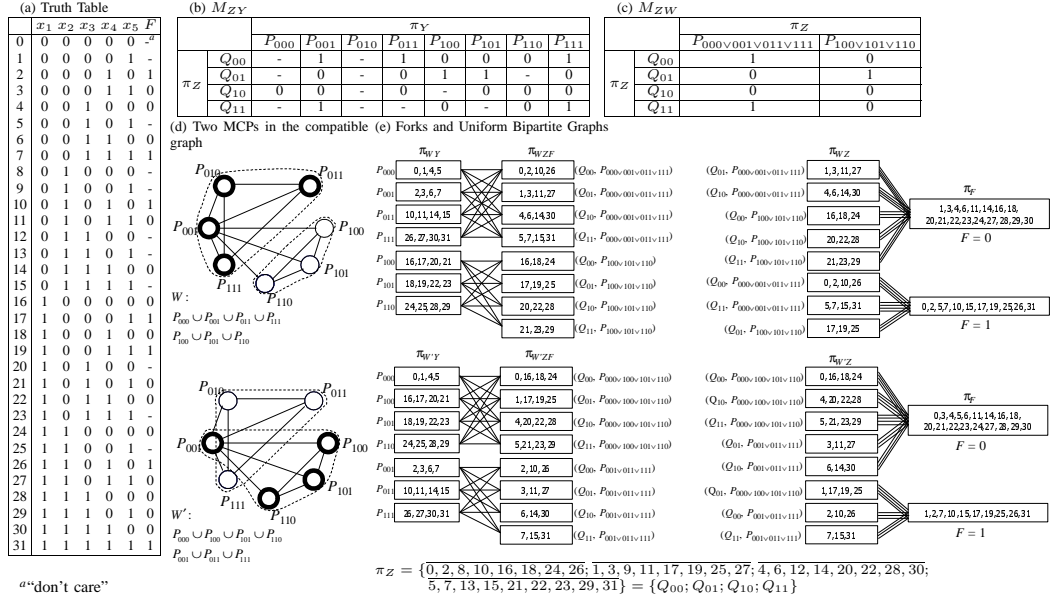

Fig. 11. Incompletely specified functional decomposition with $Y = \{x_1, x_2, x_4\}$ and $Z = \{x_3, x_5\}$.

First, the chart $M_{ZY}$ given in Fig.11(b) is established as before according to the truth table. A column that contains "don't care" entries only can be eliminated entirely because it does not provide any useful information to function $f$. For example, column $P_{010}$ in the chart $M_{ZY}$ can be discarded. Secondly, the equivalence relation between columns of $M_{ZY}$ will be relaxed to account for "don't care" entries, which give rise to a new binary relation defined as follows.

**Definition 8** *Two columns $P_i$ and $P_j$ in $M_{ZY}$ are* compatible, *if $(Q_z, P_i)$ and $(Q_z, P_j)$ have the same $F$-value or at least one of them is "don't care" for all $z$ in the free domain. A set of columns is called a* compatible column set, *if all columns in the set are pairwise compatible.*

The compatible relation among columns is not *transitive*. For instance, in the chart $M_{ZY}$, column $P_{000}$ is compatible with both column $P_{001}$, and column $P_{110}$, but $P_{001}$ and $P_{110}$ are not compatible.

The compatibility relation in the $M_{ZY}$ can be described by a *compatible graph* defined below.

(1) Each column is represented by a node.
(2) Two nodes are connected if their underlying columns are compatible.

The compatible graph of columns in the chart $M_{ZY}$ is displayed in Fig.11(d). From the above definition, a compatible column set in the chart is corresponding to a *clique* [28–30] in the compatible graph. As an example, the set of compatible columns $\{P_{000}, P_{001}, P_{011}, P_{111}\}$ forms a clique in the graph depicted in Fig.11(d).

All columns in a compatible set are equivalent, because values of their "don't care" entries can be assigned arbitrarily. Take the column set $\{P_{000}, P_{001}, P_{011}, P_{111}\}$ as an example, column $P_{011}$ is equivalent to $P_{001}$ and $P_{111}$ if $(Q_{11}, P_{011})$ is set equal to "1". It follows that all the columns in a compatible set can be merged in the same manner as discussed in S2 of FDA-$\alpha$. Hence, each block in the partition of the set of bridge variables is a clique. As a consequence, the set of bridge variables $W$ is determined by a *minimum clique partition* (MCP) of the compatible graph. The MCP is an algorithm that computes the smallest number of cliques that cover all nodes of a graph. It is known that MCP is an NP-complete problem [31].

Typically, there are multiple MCPs for a graph, and different MCPs are corresponding to different solutions. For example, Fig.11(d) shows that there exist two MCPs for the compatible graph of this example. Accordingly, there are two solutions $W$ and $W'$ for this decomposition, and their associated partitions $\pi_W$ and $\pi_{W'}$ have two blocks.

Again, theorem 1 ensures that the FD: $WZ \rightarrow F$ and MVD: $W \rightarrow\rightarrow Y$ are valid, as demonstrated by forks and uniform bipartite graphs given in Fig.11(f). Assigning "$W = 1$" and "$W = 0$" to tuples in $P_{000 \vee 001 \vee 011 \vee 111}$ and $P_{100 \vee 101 \vee 110}$ respectively, we obtain

$$W = g(x_1, x_2, x_4) = \overline{x}_1\overline{x}_2 + \overline{x}_1 x_4 + x_2 x_4,$$

and

$$F = h(W, x_3, x_5) = W\overline{x}_3\overline{x}_5 + Wx_3 x_5 + \overline{W}\,\overline{x}_3 x_5.$$

In sum, the algorithim for the decomposition of incompletely specified functions is given as follows:

**Incompletely Specified Function Decomposition Algorithm (FDA-$\delta$)**

S1. (*Initialization*) Establish the chart $M_{ZY}$ from the truth table.
S2. Eliminate columns that contain "don't care" only.
S3. Construct the compatible graph of columns in the $M_{ZY}$.
S4. Find an MCP of the compatible graph.
S5. Merge all columns in each clique of the MCP to determine blocks of $\pi_W$.

## 8 Multi-Valued Decomposition

In all examples discussed in previous sections, we assume variables of switching functions are all binary. This should by no means impose a limitation on our relational approach. In fact, tuples of multi-valued variables can be easily partitioned in the same manner as those of binary variables. Without loss of generality, all algorithms and theorems can be extended to the decomposition of *multi-valued functions*. This point is illustrated in the sequel by an example taken from [17] and given in Fig.12, where each variable possesses three values $\{lo, med, hi\}$.

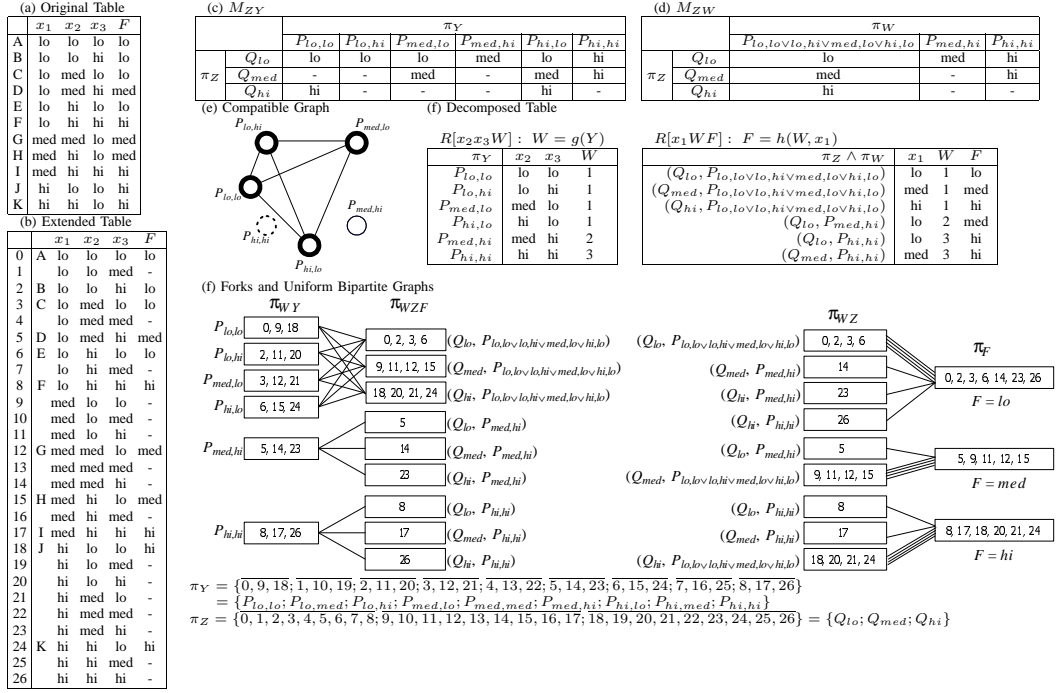

Fig. 12. Multi-valued decomposition with $Y = \{x_2, x_3\}$ and $Z = \{x_1\}$.

**Example 8** *The original table, shown in Fig.12(a), is an incompletely specified multi-valued function. Taking all X-values into account, the table is extended in Fig.12(b). Following the algorithm FDA-$\delta$ detailed in the previous section, We will described the decomposition of this multi-valued function below.*

*For given bound set $Y = \{x_2, x_3\}$ and free set $Z = \{x_1\}$, the chart $M_{ZY}$ is established according to the extended truth table. The MCP of the compatible graph shown in Fig.12(d) consists of three cliques, which give rise to three blocks of $\pi_W$, as indicated by the final chart $M_{ZW}$ shown in Fig.12(e). Assuming the bridge variable has three values, and assigning "$W = 1$", "$W = 2$", and "$W = 3$" to respective tuples in $P_{lo,lo \lor lo,hi \lor med,lo \lor hi,lo}$, $P_{med,hi}$, and $P_{hi,hi}$, the decomposed tables $R[x_2 x_3 W]$ and $R[x_1 WF]$ are provided in Fig.12(f), which agrees with those given in [17]. Finally, the uniform bipartite graphs $G(\pi_{WY} \times \pi_{WZF}, R)$ and the fork $G(\pi_{WZ} \times \pi_F, R)$ displayed in Fig.12(g) confirm that this decomposition is information-lossless.*

## 9 Conclusion

In this paper, an approach based on relational model for investigating the functional decomposition of switching function is developed. The basic concept of the decomposition of switching function lies on the notions of functional dependency and multi-valued dependency that can be portrayed by bipartite graphs with specific topological properties, which are delineated by partitions of minterms. It follows that our algorithms are procedures of constructing those specific bipartite graphs of interest to meet the lossless criteria of functional decomposition. Many design issues related to the decompositions of logic circuits, such as optimization issues, logic circuits with multiple outputs and acyclic decomposition of a large circuit into network of plural smaller circuits, are interesting and challenging research problems in their own rights, and deserve further investigations in the future.

## References

[1] H. A. Curtis, *A New Approach to the Design of Switching Circuit*. Princeton, New Jersey: D. Van Nostrand, 1962.

[2] Z. Kohavi, *Switching and finite automata theory*. New York: McGraw-Hill, 1978.

[3] T. T. Lee, T. Y. Lo, and J. Wang, "An information-lossless decomposition theory of relational information systems," *IEEE Trans. Inform. Theory*, vol. 52, no. 5, pp. 1890–1903, 2006.

[4] M. A. Perkowski and S. Grygiel, "A survey of literature on function decomposition," Portland State University, Tech. Rep. Ver. IV, 1995.

[5] R. L. Ashenhurst, "The decomposition of switching functions," Bell Laboratories, Tech. Rep. BL-1(11), 1952.

[6] J. P. Roth, "Minimization over boolean trees," *IBM Journal of Research and Development*, vol. 4, no. 5, pp. 543–558, Nov. 1960.

[7] R. M. Karp *et al.*, "A computer program for the synthesis of combinational switching circuits," in *Proc. AIEE Annual Symposium on Switching Circuit Theory and Logical Design*, 1961, pp. 182–194.

[8] R. Brayton and C. McMullen, "The decomposition and factorization of boolean expressions," in *Proc. of the International Symposium on Circuits and Systems*, 1982, pp. 49–54.

[9] R. Murgai *et al.*, "Logic synthesis for programmable gate arrays," in *Proc. 27th ACM/IEEE conference on Design automation*, 1990, pp. 620–625.

[10] M. A. Perkowski, M. Helliwell, and P. Wu, "A unified approach to designs with multiplexers and to the decomposition of boolean functions," in *Proc. ASEE Annual Conference*, 1988, pp. 1610–1619.

[11] H. Selvaraj *et al.*, "A generalized decomposition of boolean functions and its application in FPGA-based synthesis," in *IFIP Workshop on Logic and Architecture Synthesis*, 1993, pp. 147–166.

[12] L. Józwiak and A. Chojnacki, "Effective and efficient FPGA synthesis through general functional decomposition," *Journal of Systems Architecture*, vol. 49, no. 4-6, pp. 247–265, 2003.

[13] K. M. Waliuzzaman and Z. G. Vranesic, "On decomposition of multiple-valued switching functions," *Computer Journal*, vol. 13, pp. 359–362, 1970.

[14] S. L. Hurst, "Multiple-valued logic – its status and its future," *IEEE Trans. Comput.*, vol. 33, no. 12, pp. 1160–1179, Dec. 1984.

[15] W. C. Kabat and A. S. Wojcik, "Automated synthesis of combinational logic using theorem-proving techniques," *IEEE Trans. Comput.*, vol. 34, no. 7, pp. 610–632, July 1985.

[16] K. Y. Fang and A. S. Wojcik, "Modular decomposition of combinational multiple-valued circuits," *IEEE Trans. Comput.*, vol. 37, no. 10, pp. 1293–1301, Oct. 1988.

[17] B. Zupan *et al.*, "Machine learning by function decomposition," in *Proc. 14th International Conference on Machine Learning*, 1997, pp. 421–429.

[18] C. Files and M. Perkowski, "Multi-valued functional decomposition as a machine learning method," in *Proc. of the IEEE 28th International Symposium on Multiple-Valued Logic*, 1998, pp. 173–178.

[19] R. K. Brayton *et al.*, *Logic Minimization Algorithms for VLSI Synthesis*. Norwell: Kluwer Academic, 1984.

[20] E. F. Codd, "A relational model of data for large shared data banks," *Communications of ACM*, vol. 13, no. 6, pp. 377–387, June 1970.

[21] D. Maier, *The Theory of Relational Databases*. Rockville, Maryland: Computer Science Press, 1983.

[22] J. Hartmanis and R. E. Sterns, *Algebraic structure theory of sequential machines*. Englewood Cliffs: NJ: Prentice-Hall, 1966.

[23] G. Birkhoff, *Lattic Theory*, 2nd ed. Providence, RI: AMS, 1948, vol. 25, American Mathematical Society Colloquium Publications.

[24] G. C. Rota, "The many lives of lattice theory," *Notices American Mathematics Society*, vol. 44, no. 11, pp. 1440–1445, 1997.

[25] T. T. Lee, "An algebraic theory of relational databases," *The Bell System Technical Journal*, vol. 62, no. 10, pp. 3159–3204, Dec. 1983.

[26] F. P. Preparata and R. T. Yeh, *Introduction to Discrete Structures*. Boston: Addison-Wesley, 1973.

[27] R. J. Wilson, *Introduction to graph theory*. New York: Academic Press, 1973.

[28] C. Bron and J. Kerbosch, "Finding all cliques of an undirected graph," *Communications of ACM*, vol. 16, no. 9, pp. 575–577, Sept. 1973.

[29] R. Carraghan and P. M. Pardalos, "An exact algorithm for the maximum clique problem," *Operations Research Letters*, vol. 9, no. 6, pp. 375–382, Nov. 1999.

[30] P. R. J. Östergård, "A fast algorithm for the maximum clique problem," *Discrete Applied Mathematics*, vol. 120, no. 1, pp. 197–207, Aug. 2002.

[31] M. Garey and D. Johnson, *Computers and intractibility: a guide to the theory of NP-Completeness*. San Fransico: W. H. Freeman and Company, 1979.